\documentclass[manuscript]{aastex}

\shorttitle{Magnetically Torqued NDAF}
\shortauthors{Lei et al.}

\begin{document}
\title{Magnetically Torqued Neutrino-Dominated Accretion Flows for
Gamma-ray Bursts}

\author{W. H. Lei\altaffilmark{1}, D. X. Wang\altaffilmark{1,2}, L. Zhang\altaffilmark{1},Z. M. Gan\altaffilmark{1},Y. C. Zou\altaffilmark{1}}
\affil{School of Physics, Huazhong University of Science and
Technology, Wuhan, 430074, China}

\and

\author{Y. Xie\altaffilmark{3}}

\affil{School of Physics and Information Engineering, Shanxi Normal
University, Linfen, 041004, China}

\altaffiltext{1}{School of Physics, Huazhong University of Science
and Technology, Wuhan, 430074, China}

\altaffiltext{2}{Send offprint requests to: D. X. Wang
(dxwang@mail.hust.edu.cn)}

\altaffiltext{3}{School of Physics and Information Engineering,
Shanxi Normal University, Linfen, 041004, China}

\begin{abstract}

Recent observations and theoretical work on gamma-ray bursts (GRBs)
favor the central engine model of a Kerr black hole (BH) surrounded
by a magnetized neutrino-dominated accretion flow (NDAF). The
magnetic coupling between the BH and disk through a large-scale
closed magnetic field exerts a torque on the disk, and transports
the rotational energy from the BH to the disk. We investigate the
properties of the NDAF with this magnetic torque. For a rapid
spinning BH, the magnetic torque transfers enormous rotational
energy from BH into the inner disk. There are two consequences: (i)
the luminosity of neutrino annihilation is greatly augmented; (ii)
the disk becomes thermally and viscously unstable in the inner
region, and behaves S-Shape of the surface density versus accretion
rate. It turns out that magnetically torqued NDAF can be invoked to
interpret the variability of gamma-ray luminosity. In addition, we
discuss the possibility of restarting the central engine to produce
the X-ray flares with required energy.
\end{abstract}


\section{INTRODUCTION}

The nature of the central engine of gamma-ray bursts (GRBs) remains
unclear. Currently favored models invoke a binary merger or a
collapse of compact objects. These models lead to the formation of a
transient hot and dense accretion torus/disk around a black hole
(BH) of a few solar masses.

The typical mass accretion rates in GRB models are extremely high,
of the order of a fraction of solar mass up to a few solar masses
per second. Under such conditions, the disk becomes dense and hot
enough in the inner regions to cool via neutrino losses. For this
reason, Popham, Woosley {\&} Fryer (1999, hereafter PWF99) named
these disk neutrino-dominated accretion flows (NDAF). Energy
extraction from the BH-accretion disk for powering GRB is also
possible, such as by Blandford-Znajek (BZ) process (see Lee et al.
2000 for a review of this model), and Blandford-Payne (BP) process
and Parker instabilities in the disk (Narayan et al. 1992; Meszaros
{\&} Rees 1997).

NDAF has been extensively discussed by many authors, e.g., PWF99,
Narayan, Piran {\&} Kumar (2001, hereafter NPK01), Kohri {\&}
Mineshige (2002), Di Matteo, Perna {\&} Narayan (2002, hereafter
DPN), Chen {\&} Beloborodov (2006), Janiuk et al. (2004, 2007) and
Gu, Liu \& Lu (2006, hereafter GLL06). However, due to its low
energy conversion efficiencies and the effects of neutrino opacity,
the power produced by neutrino-anineutrino annihilation can hardly
to match those of the energetic short-hard GRBs (e.g. GRB080913) and
X-ray flares. Recently, some authors (e.g., DPN02, Fan et al. 2005,
Shibata et al. 2006 and 2007 and Perez-Ramirez et al. 2008)
suggested that the MHD process should be considered in the disk
model. Moreover, as shown in Shibata et al. (2006, 2007), the
magnetic braking and magneto-rotational instability (MRI, Balbus
{\&} Hawley 1991) in the disk play a role in angular momentum
transporting, which causes turbulent motion, resultant shock
heating, and mass accretion onto the BH. On the other hand,
researches showed that the magnetic fields can be magnified up to
$10^{15} \sim 10^{16}$G by virtue of MRI or dynamo process (Pudritz
{\&} Fahlman 1982 and references therein) in hyperaccretion disk.
These considerations stimulate us to discuss the magnetized NDAF.

Based on the work of NPK01 and DPN02, Xie et al. (2007) discussed
the BZ and BP processes in NDAF. They found that the jet of GRB may
be magnetically-dominated, which is also obtained by MHD simulations
of Mizuno et al.(2004).

Recently, the magnetic coupling (MC) between the central spinning BH
and their surrounding accretion disk has been paid much attention
(e.g. Blandford 1999; van Putten 1999; Li {\&} Paczynski 2000; Li
2002; Wang et al. 2002). As a variant of the BZ process, the MC
process exerts a torque on the disk, and transports the rotational
energy from the BH to the disk. The effects of MC torque has been
discussed in some disk models, for example, Lai (1998) and Lee
(1999) in a neutron star with slim disk, Li (2002), Wang et al.
(2002, 2003), Kluzniak and Rappaport (2007) in a compact object with
thin disk, Ye et al. (2007) and Ma et al. (2007) in a BH with
advection-dominated accretion flow (ADAF). It is found, the disk
properties are greatly changed and its luminosity is augmented
significantly due to the rotational energy of BH extracted in the MC
process. Therefore, it is attractive for us to investigate the
effects of MC torque on NDAF. To highlight the effects of MC torque,
we ignore other MHD process, such as BZ and BP mechanism, and we
refer to this model as MCNDAF.

This paper is organized as follows. In Sect. 2 we describe the
MCNDAF model, which is a relativistic steady state thin disk. The
effects of MHD stress are described by the dimensionless parameter
$\alpha $. The main equations are based on DPN02 and NPK01.
Recently, GLL06, Chen {\&} Beloborodov (2006) and Shibata et al.
(2006, 2007) argued that the general relativistic (GR) effects are
important for NDAF, so we introduce GR correction factors to the
equations. The MC torque appears in the angular momentum equation.
We solve the set of equations for the solutions in MCNDAF in Sect.
3, and compute the neutrino and neutrino annihilation luminosities.
Following GLL06, we include the neutrino radiation from the
optically thick region in the computation for the neutrino
luminosity. To show the effects of MC torque, we compare it with
previous results. In Sect. 4 we discuss the stability of the
accretion flow. We also discuss the physical origin of the
instabilities in MCNDAF. Finally, we summarize our results and
discuss some related issues in Sect. 5.

\section{NEUTRINO-DOMINATED ACCRETION FLOWS WITH MC EFFECTS}

Considering that central BHs are rapidly rotating in most candidate
GRB engines, we discuss a model of a steady state disk around a Kerr
BH, in which neutrino loss and transfer are taken into account. Our
model is presented in the context given by DPN02, and the GR
corrections are adopted from Riffert {\&} Herold (1995).

As mentioned by DPN02 and PWF99, although in GRB central engines the
accretion rate may vary, it is expected to vary significantly only
in the outer disk. Hence it seems reasonable to study the main
properties in the inner neutrino-cooled disk by assuming a constant
accretion rate.

Because the gas cools efficiently, we are entitled to discuss the
MCNDAF model in the context of a thin disk (Shakura {\&} Sunyaev
1973). The accuracy of the thin-disk approximation is not perfect at
large radii, where the disk is thick. Fortunately, the details of
the outer region have little effect on the solution for the
neutrino-cooled disk (Chen {\&} Beloborodov 2006).

The MCNDAF model is a relativistic steady thin disk, and the
large-scale magnetic field contributing to the MC process and the
small-scale tangled magnetic field related to the viscosity are
included. We assume that these two kinds of fields work
independently, and the large-scale magnetic field remains constant
at the BH horizon. Following Blandford (1976) we assume that the
magnetic field $B_z $ on the disk varies as $B_z { \propto} \xi ^{ -
n} $, where $\xi \equiv r / r_{ms} $ is the disk radius in terms of
the marginally stable orbit $r_{ms} $ (Novikov {\&} Thorne 1973),
and $n$ is the power law index indicating the degree of
concentration of the magnetic field in the central region of the
disk.

Based on equipartition relation the magnetic field at the horizon is related
to the mass density at the inner disk as follows (McKinney 2005),

\begin{equation}
\label{eq1}
\frac{B_H^2 }{8\pi } = \rho _{0,disk} c^2,
\end{equation}

\noindent
where $\rho _{0,disk} \equiv \dot {M}t_g / r_g^3 $, $t_g = GM / c^3$ and
$r_g = GM / c^2$.

The MC torque is derived in Wang et al. (2002) based on an
equivalent circuit given by Macdonald {\&} Thorne (1982) as follows,

\begin{equation}
\label{eq2}
T_{MC} / T_0 = 4a_\ast (1 + q)\int_0^{\pi / 2} {\frac{(1 - \beta )\sin
^3\theta d\theta }{2 - (1 - q)\sin ^2\theta }}
\end{equation}

\noindent where $T_0 \approx 3.26\times 10^{45}(\textstyle{{B_H }
\over {10^{15}G}})^2(\textstyle{M \over {M_ \odot }})^3g \cdot cm^2
\cdot s^{ - 2}$, $a_* \equiv J c/(G M^2)$ is the dimensionless BH
spin parameter defined by the BH mass $M$ and angular momentum $J$,
$q = \sqrt {1 - a_\ast ^2 } $ and $\beta \equiv \Omega_D/ \Omega_H$
is the ratio of the angular velocity of the disk $\Omega_D =
((r^3/GM)^{1/2}+a_* GM/c^3)^{-1}$ to that of the horizon
$\Omega_H=a_* c^3/[2 G M (1+q)]$.

The mapping relation between the angular coordinate $\theta $ on the horizon
and the radial coordinate $\xi $ on the disk is derived based on the
conservation of magnetic flux as follows (Wang et al. 2003),

\begin{equation}
\label{eq3}
\cos \theta = \int_1^\xi {\Theta (a_\ast ;\xi ,n)d\xi }
\end{equation}

\noindent where $\xi=(r/r_{ms})^{1/2}$,

\begin{equation}
\label{eq4}
\Theta (a_\ast ;\xi ,n) = \frac{\xi ^{1 - n}\chi _{ms}^2 \sqrt {1 + a_\ast
^2 \chi _{ms}^{ - 4} \xi ^{ - 2} + 2a_\ast ^2 \chi _{ms}^{ - 6} \xi ^{ - 3}}
}{2\sqrt {(1 + a_\ast ^2 \chi _{ms}^{ - 4} + 2a_\ast ^2 \chi _{ms}^{ - 6}
)(1 - 2\chi _{ms}^{ - 2} \xi ^{ - 1} + a_\ast ^2 \chi _{ms}^{ - 4} \xi ^{ -
2})} }.
\end{equation}

\noindent where $\chi_{ms}=(r_{ms}/r_{g})^{1/2}$.

A number of works (PWF99; Chen {\&} Beloborodov, 2006) used accurate
equations of relativistic hydrodynamics in Kerr spacetime to study
NDAF. GLL06 found that the GR effect must be taken into account for
the power of GRB.

The relativistic correction factors for a thin accretion disk around a Kerr
BH have been given by Riffert {\&} Herold (1995),

\begin{equation}
\label{eq5}
A = 1 - \frac{2GM}{c^2r} + (\frac{GMa_\ast }{c^2r})^2,
\end{equation}

\begin{equation}
\label{eq6}
B = 1 - \frac{3GM}{c^2r} + 2a_\ast (\frac{GM}{c^2r})^{3 / 2},
\end{equation}

\begin{equation}
\label{eq7}
C = 1 - 4a_\ast (\frac{GM}{c^2r})^{3 / 2} + 3(\frac{GMa_\ast
}{c^2r})^2,
\end{equation}

\begin{equation}
\label{eq8}
D = \int_{r_{ms} }^r {\frac{\textstyle{{x^2c^4} \over {8G^2}} -
\textstyle{{3xMc^2} \over {4G}} + \sqrt {\textstyle{{a_\ast ^2 M^3c^2x}
\over G}} - \textstyle{{3a_\ast ^2 M^2} \over 8}}{\textstyle{{\sqrt {rx} }
\over 4}(\textstyle{{x^2c^4} \over {G^2}} - \textstyle{{3xMc^2} \over {G}}
+ 2\sqrt {\textstyle{{a_\ast ^2 M^3c^2x} \over G}} )}dx} .
\end{equation}

The equation of the conservation of mass remains valid, while
hydrostatic equilibrium in the vertical direction leads to a
corrected expression for the half thickness of the disk (Riffert
{\&} Herold 1995; Reynoso, Romero {\&} Sampayo 2006):

\begin{equation}
\label{eq9}
H \simeq \sqrt {\frac{Pr^3}{\rho GM}} \sqrt {\frac{B}{C}} .
\end{equation}

The viscous shear $t_{r\varphi } $ is corrected by

\begin{equation}
\label{eq10}
t_{r\varphi } = - \alpha P\sqrt {\frac{A}{BC}} .
\end{equation}

The basic equations of MCNDAF are given as follows.

1. The continuity equation:

\begin{equation}
\label{eq11}
\dot {M} = - 2\pi rv_r \Sigma .
\end{equation}

2. The total pressure consists of five terms, radiation pressure,
gas pressure, degeneracy pressure, neutrino pressure and magnetic
pressure:

\begin{equation}
\label{eq12}
P = \frac{11}{12}aT^4 + \frac{\rho kT}{m_p }(\frac{1 + 3X_{nuc} }{4}) +
\frac{2\pi hc}{3}(\frac{3}{8\pi m_p })^{4 / 3}(\frac{\rho }{\mu _e })^{4 /
3} + \frac{u_\nu }{3} + P_{mag} .
\end{equation}

\noindent where $P_{mag} = \beta _t P$ is the magnetic pressure
contributed by the tangled magnetic field in the disk, and $\beta _t
$ is the ratio of the magnetic pressure to the total pressure.
$u_\nu $ is the neutrino energy density defined as (Popham {\&}
Narayan 1995)

\begin{equation}
\label{eq13}
u_\nu = (7 / 8)aT^4\sum {\frac{\tau _{\nu _i } / 2 + 1 / \sqrt 3 }{\tau
_{\nu _i } / 2 + 1 / \sqrt 3 + 1 / (3\tau _{a,\nu _i } )}}
\end{equation}

In equation (\ref{eq13}) $\tau _{\nu _i } = \tau _{a,\nu _i } + \tau _{s,\nu _i } $
is the sum of absorptive and scattering optical depths calculated for each
neutrino flavor $(\nu _e ,\nu _\mu ,\nu _\tau )$. The absorptive optical
depths for the three neutrino flavors are (Kohri et al. 2005)

\begin{equation}
\label{eq14}
\tau _{a,\nu _e } \simeq 2.5\times 10^{ - 7}T_{11}^5 H + 4.5\times 10^{ -
7}T_{11}^2 X_{nuc} \rho _{10} H,
\end{equation}

\begin{equation}
\label{eq15}
\tau _{a,\nu _\mu } = \tau _{a,\nu _\tau } \simeq 2.5\times 10^{ -
7}T_{11}^5 H,
\end{equation}

\noindent where $X_{nuc} $ is the mass fraction of free nucleons
approximately given by (e.g., PWF99; Qian {\&} Woosley 1996),

\begin{equation}
\label{eq16}
X_{nuc} \simeq 34.8\rho _{10}^{ - 3 / 4} T_{11}^{9 / 8} \exp ( - 0.61 /
T_{11} ).
\end{equation}

The total scattering optical depth is given by DPN02 as

\begin{equation}
\label{eq17}
\tau _{s,\nu _i } \simeq 2.7\times 10^{ - 7}T_{11}^2 \rho _{10} H.
\end{equation}

3. Combining the conservation of the angular momentum with equation
(\ref{eq11}), we have

\begin{equation}
\label{eq18}
\frac{d}{dr}(\dot {M}l) + 4\pi rH_{MC} = \frac{d}{dr}g = - \frac{d}{dr}(4\pi
r^2t_{r\varphi } H),
\end{equation}

\noindent
where $l$ is the specific angular momentum of the accreting gas. The flux of
angular momentum transferred magnetically from the BH to the disk, $H_{MC}
$, is related to the MC torque $T_{MC} $ by

\begin{equation}
\label{eq19}
T_{MC} = 4\pi \int_{r_{ms} }^r {H_{MC} rdr} .
\end{equation}

Vanishing of $t_{r\varphi } $ (or $g)$ at $r_{ms} $ leads to

\begin{equation}
\label{eq20}
\dot {M}r^2\sqrt {\frac{GM}{r^3}} \frac{D}{A} + T_{MC} = g = - 4\pi
r^2t_{r\varphi } H = 4\pi r^2H\alpha P\sqrt {\frac{A}{BC}}
\end{equation}

4. The equation for the energy balance is

\begin{equation}
\label{eq21}
Q^ + = Q^ -
\end{equation}

\noindent where $Q^ + = Q_{vis} $ represents the viscous
dissipation, and $Q^ - = Q_\nu + Q_{photo} + Q_{adv} $ is the total
cooling rate due to neutrino losses $Q_\nu $, photodisintegration
$Q_{photo} $ and advection $Q_{adv} $. We employ a bridging formula
for calculating $Q_\nu $, which is valid in both the optically thin
and thick cases. The expressions for $Q_\nu $, $Q_{photo} $ and
$Q_{adv} $ are (DPN02; GLL06)

\begin{equation}
\label{eq22}
Q_\nu = \sum {\frac{(7 / 8\sigma T^4)}{(3 / 4)(\tau _{\nu _i } / 2 + 1 /
\sqrt 3 + 1 / (3\tau _{a,\nu _i } ))}} ,
\end{equation}

\begin{equation}
\label{eq23}
Q_{photo} = 10^{29}\rho _{10} v_r \frac{dX_{nuc} }{dr}H\mbox{ }erg \cdot
cm^{ - 2}s^{ - 1},
\end{equation}

\begin{equation}
\label{eq24}
Q_{adv} \simeq v_r \frac{H}{r}(\frac{11}{3}aT^4 + \frac{3}{2}\frac{\rho
kT}{m_p }\frac{1 + X_{nuc} }{4} + \frac{4u_\nu }{3}),
\end{equation}

\noindent where $4u_\nu / 3$ is the entropy density of neutrinos.
Note that the cooling function given by the bridging formula reduces
to the optically thin expression for small optical depths (as
adopted in PWF99) but differs significantly from the latter at
optical depths $\sim $ 1.

By considering the MC effects, the heating rate $Q_{vis} $ is expressed as

\begin{equation}
\label{eq25} Q_{vis} = - \frac{g{\Omega }'_D }{4\pi r} =
\frac{3GM\dot {M}}{8\pi r^3}\frac{D}{B} - \frac{T_{MC} {\Omega
}'_D}{4\pi r},
\end{equation}

\noindent
where the second term is the MC contribution.

As we can see from equation (\ref{eq20}), the magnetic torque may deposit angular
momentum in the inner disk, and this extra angular momentum must be
transported outwards by the viscous torque in the disk, resulting in energy
dissipation and increasing the disk luminosity based on equation (\ref{eq25}).

Defining $Q_G = 3GM\dot {M}D / (8\pi r^3B)$ and $Q_{MC} = - T_{MC} {\Omega
}'_D / (4\pi r)$ as the contributions due to the gravitational release and
the MC process, respectively, we have the ratio $\eta \equiv Q_{MC} / Q_G $
versus the disk radius $R \equiv r \mathord{\left/ {\vphantom {r {r_g }}}
\right. \kern-\nulldelimiterspace} {r_g }$ as shown in Figure 1.

From Figure 1 we find that $Q_{MC} $ is much greater than $Q_G $ in
the inner disk, where the neutrino cooling dominates. The ratio
$\eta $ is very sensitive to the value of $a_\ast $ and $n$, it
increases monotonically with the increasing $a_\ast $ and $n$. This
implies that the MC effects are more important for the greater
$a_\ast $ and $n$. For simplicity, we choose $a_\ast = 0.9$ and
$n$=3 in the calculations, and discuss the influence of their values
in Sect. 5.

We solve numerically equations (\ref{eq12}), (\ref{eq20}) and (\ref{eq21}) to find the disk
temperature $T$ and density $\rho $ versus the disk radius with the given
$a_\ast $, $n$ and $\dot {m}$ (where $\dot {m}$ is the accretion rate in units
of $M_ \odot s^{ - 1})$. We take $X_{nuc} = 1$ for the fully
photodisintegrated nuclear, which is appropriate in the inner disk. In the
calculation, we do not include the cooling term arising from the
photodisintegration, because it is much less than the neutrino cooling rate
in the inner disk (Janiuk et al. 2004). Furthermore, $\alpha = 0.1$, $M =
7M_ \odot $ and $\beta _t = 0.1$ are adopted in calculations.

\section{EFFECTS OF THE MC TORQUE ON NEUTRINO ANNIHILATION LUMINOSITY}

As discussed in Sect. 2, the MC process applies a strong torque on the disk,
resulting in huge viscous dissipation. This would lead to a more powerful
neutrino radiation, and neutrino annihilation luminosity. Here, we will show
the effects of MC torque on the neutrino annihilation luminosity.

To show this, we compare the results of MCNDAF with the NDAF model
without MC (hereafter NDAF refers to the model without MC). GLL06
pointed out that the GR effects and the neutrino radiation from the
optically thick region are important for the NDAF luminosity.
Therefore we include these two effects in our calculations for both
MCNDAF and NDAF.

The neutrino luminosity from the accretion flow is

\begin{equation}
\label{eq26}
L_\nu = 4\pi \int_{r_{ms} }^{r_{out} } {Q_\nu rdr} .
\end{equation}

We are interested primarily in the properties of the inner accretion
flow, where neutrino processes are important. As argued in PWF99,
NPK01 and DPN02, the flows are fully advection-dominated for $r >
100r_g $, since neutrino cooling is not important and photons are
completely trapped. Thus we concentrate the discussion in the region
from $r_{ms} $ to $r_{\max } = 100r_g $.

Our method for calculating neutrino annihilation is similar to PWF99
and Rosswog et al. (2003). The disk is modeled as a grid of cells in
the equatorial plane. A cell $k$ has its neutrino mean energy
$\varepsilon _{\nu _i }^k $ and luminosity $l_{\nu _i }^k $, and the
height above (or below) the disk is $d_k $. The angle at which
neutrinos from cell $k$ encounter antineutrinos from another cell
$k'$ at that point is denoted as $\theta _{k{k}'} $. Then the
neutrino annihilation luminosity at that point is given by the
summation over all pairs of cells,

\begin{equation}
l_{\nu \bar{\nu}}  =  A_1 \sum_k \frac{l^k_{\nu_i}}{d_k^2} \sum_{k'}
\frac{l^k_{\nu_i}}{d_k^2}
(\epsilon^k_{\nu_i}+\epsilon^{k'}_{\bar{\nu}_i})(1-cos\theta_{kk'})^2
 + A_2 \sum_k \frac{l^k_{\nu_i}}{d_k^2} \sum_{k'} \frac{l^k_{\nu_i}}{d_k^2} \frac{\epsilon^k_{\nu_i}+\epsilon^{k'}_{\bar{\nu}_i}}{\epsilon^k_{\nu_i} \epsilon^{k'}_{\bar{\nu}_i}}(1-cos\theta_{kk'})
\end{equation}

\noindent
where $A_1 \approx 1.7\times 10^{ - 44}cm \cdot ergs^{ - 2} \cdot s^{ - 1}$
and $A_2 \approx 1.6\times 10^{ - 56}cm \cdot ergs^{ - 2}s^{ - 1}$.

The total neutrino annihilation luminosity is obtained by integrating over
the whole space outside the BH and the disk,

\begin{equation}
\label{eq27}
L_{\nu \bar {\nu }} = 4\pi \int\!\!\!\int {l_{\nu \bar {\nu }} rdrdz}
\end{equation}

As shown in Figure 2, the variations of $L_\nu $ and $L_{\nu \bar
{\nu }} $ versus $\dot {m}$ for MCNDAF are indicated by the thin and
thick solid lines, respectively, while those for NDAF are marked by
the dotted and dashed lines, respectively. It is found that $L_\nu $
and $L_{\nu \bar {\nu }} $ are greatly strengthened in MCNDAF. This
means the spin energy acts as a powerful source for NDAF.

According to our calculations for MCNDAF $L_{\nu \bar {\nu }} $
varies from $3.7\times 10^{49}ergs \cdot s^{ - 1}$ to $1.4\times
10^{54}ergs \cdot s^{ - 1}$ for $0.01 < \dot {m} < 10$. We find that
$L_{\nu \bar {\nu }} $ nearly stays constant around $\sim
10^{54}ergs \cdot s^{ - 1}$ for the accretion rate above $\dot
{m}\sim 0.5$. This implies that the effect of neutrino optical depth
becomes important. Our results for NDAF are in good agreement with
those given by PWF99 and GLL06, but larger than those in DPN02. This
is because the GR effects are taken into account in this paper as
well as in PWF99 and GLL06.

\section{STABILITY ANALYSIS: THERMAL-VISCOUS INSTABILITY}

DPN02 discussed the thermal, viscous and gravitational stability
properties of NDAF solutions. They found that NDAF is stable in most
cases. But this result is not consistent with the variability in GRB
lightcurve. To explain the X-ray flares, it is need that after the
prompt gamma-ray emission has ceased, the central engine can be
restarted (Fan {\&} Wei 2005; Zhang et al. 2006). Based on this
scenario, Perna et al. (2006) suggested that the X-ray flares could
be produced by accretion of matter after the breaking of the disk
due to the setting up of various instabilities either gravitational
or viscous. Therefore, it is attractive for us to examine whether
MCNDAF solution is stable.

The condition for viscous stability is given by

\begin{equation}
\label{eq28}
\frac{d\dot {M}}{d\Sigma } > 0,
\end{equation}

The stability curves for several radii in the disk are shown in Figure 3.

From Figure 3, we find the $\dot {m} - \Sigma $ curves show S-Shape,
in which the branch of solutions with negative slop is viscously
unstable. It is shown that the viscous instability occurs at larger
radius for larger accretion rate. However, we find this unstable can
only occur at $\dot {m} > 0.086$, see Figure 5.

It is clearly shown in Figure 3 that the disk is unstable at
$\dot{m}=0.5$ for $R=3$, $\dot{m}=2$ for $R=10$, and $\dot{m}=10$
for $R=20$. To understood this S-Shape we draw Figure 4.

Inspecting Figures 3 and 4, we find the viscous instability occurs
when the disk is neutrino cooing and radiation pressure dominated.
For $\dot {m} < 0.086$, the MC torque will become very small, and
the disk is optically thin to neutrinos. As discussed in NPK01, an
optically thin NDAF is viscously stable for all pressure cases.
Therefore, we will find no viscous instability when $\dot {m} <
0.086$. If the accretion rate beyond $0.086M_ \odot s^{ - 1}$, the
MC torque transport enormous energy into the inner disk, and make
the inner disk optically thick to neutrinos. In this time, we have
$\dot {m} \propto \Sigma ^{ - 1}$ for neutrino cooling and radiation
pressure dominated case, and the disk will be viscously unstable. If
gas pressure dominates, we have $\dot {m} \propto \Sigma $ and $\dot
{m} \propto \Sigma ^3$ for optically thin and thick NDAF,
respectively. In the region where advection dominated, the disk is
viscous stable, which is the well-known property of slim disk.

From the above analysis, we conclude that the appearance of the viscous
instability is due to the MC torque. First, the MC torque results in an
optically thick neutrino-cooling dominated flow in the inner disk part for
high accretion rates. Second, the high opacity will drop the
neutrino-cooling rate, and leave a hot and thick disk. In this region, the
gas pressure drops and radiation pressure becomes important. We checked that the unstable solutions appear for black hole spin $a_*>0.35$ and magnetic pressure $\beta_t< 0.8$.

The disk is thermally unstable if $(d\ln Q^ + / d\ln T)\left|
{_\Sigma } \right. > (d\ln Q^ - / d\ln T)\left| {_\Sigma } \right.$.
Then any small increase (decrease) in temperature leads to heating
rate which is more (less) than the cooling rate, and as a
consequence a further increase (decrease) of the temperate. For an
optically thick NDAF in which neutrino cooling and radiation
pressure dominates, we have $Q^+ \propto T^8 /\Sigma$ and $Q^-
\propto T^4$. Therefore, the disk is also thermally unstable at the
negative slope of the S-Shape $\dot {m} - \Sigma $ curves shown in
Figure 3.

It is noticed that Janiuk et al. (2007) obtain the viscous instability
occurring at $\dot {m} > 10$ without the S-Shape curves, which is caused by
the behavior distribution for $X_{nuc} $ and photodisintegration term.
However, in MCNDAF, the photodisintegration term is not included and
$X_{nuc} = 1$ is assumed for simplicity. Therefore, we infer that the
S-Shape curves in MCNDAF arise from the MC effects.

Finally, we check the gravitational stability condition, for which the
Toomre parameter $Q_T $ should be larger than unity. For Keplerian disk,
$Q_T $ is given by $Q_T = c_s \kappa / (\pi G\Sigma ) = \Omega _D^2 / (\pi
G\rho )$. $Q_T $ decrease with increasing $r$ so that the flow is most unstable
on the outside.

\section{SUMMARY AND DISCUSSION}

In this paper we investigate some properties in MCNDAF. The angular momentum
deposited in the disk by the magnetic torque exerted by the BH leads to a
substantial additional dissipation of energy in the disk, which is greater
than that expected from gravitational release alone. Therefore, we obtain a
series of MCNDAF solutions being different significantly from NDAF.

The main results are summarized as follows.

1. The neutrino annihilation luminosity in MCNDAF varies from
$3.7\times 10^{49}ergs \cdot s^{ - 1}$ to $1.4\times 10^{54}ergs
\cdot s^{ - 1}$ for $0.01 < \dot {m} < 10$, while for NDAF the value
range is from $1.2\times 10^{45}ergs \cdot s^{ - 1}$ to $2.6\times
10^{53}ergs \cdot s^{ - 1}$, i.e., it is greatly improved by the MC
torque.

Recently, observations show that half of the Swift bursts exhibit
X-ray flares. Fan et al. (2005) pointed out that the energy from
NDAF cannot match the X-ray flares detected in GRB 050724 of $\sim
$100 s, which is also the time scale of the central engine. The time
averaged isotropic luminosity of the X-ray flare component is $L_X
\sim 10^{48}ergs \cdot s^{ - 1}$. If we assume that the total mass
available for accretion is $\sim 1M_ \odot $ (a typical value for
the compact object merger scenarios and massive star collapse
scenario), and that most of the mass is accreted during the X-ray
flare phase, the time averaged accretion rate is about $0.01M_ \odot
s^{ - 1}$. At this accretion rate, the jet luminosity powered by
neutrino annihilation is $L_{\nu \bar {\nu }} \sim 10^{45}ergs \cdot
s^{ - 1}$ for NDAF without MC torque, and it is insufficient to
power the X-ray flares. But the power produce by MCNDAF can satisfy
this requirement.

Very recently, the new observation of the highest redshift (z=6.7)
swift source GRB080913 puts a very strong constraint on the central
engine (Perez-Ramirez et al. 2008). The duration of this short burst
is $T_{90} = 8s$, and the isotropic energy required is $E_{iso}
\approx 7\times 10^{52}ergs$. If the central engine is NDAF without
MC torque, the jet collimation factor obeys $f_\Omega < 7\times 10^{
- 4}$, i.e., the jet should be strongly collimated. If invoke the MC
torque in NDAF, the observed energy can be easily satisfied.

2. The disk becomes thermally and viscously unstable in its inner
region for $\dot {m} > 0.086$. It is very interesting for us to
obtain the $\dot {m} - \Sigma $ curves behaving S-Shape, which may
produce a limit-cycle activity. The disk is rather thick in the
inner region, and therefore the thermal and viscous timescales are
close to each other. The timescale for the unstable can be estimated
by the viscous timescale,$t_{vis} = [1 / (\alpha \Omega )](r /
H)^2$, which is about 10ms at the inner disk. Following the
discussions in Janiuk et al. (2007), these instabilities will lead
to a variable energy output on millisecond timescales, which may
correspond to the variability in the gamma-ray luminosity. The
irregularity in the overall outflow can also help produce internal
shocks. On the other hand, the thermal-viscous instability may be
accompanied by the disk breaking, which can lead to the several
episodic accretion events and explain the long-time activity
accounting for the X-ray flares.

Therefore, the MCNDAF can easily power both GRBs and its X-ray flares,
naturally interpret the variability in the gamma-ray luminosity, and explain
the production of X-ray flares. However, there are several issues should be
addressed.

First, in all of the MCNDAF solutions, we assume $n$=3 and a large
BH spin $a_\ast = 0.9$ but does not give any reason. From Figure 1
we find the MC contribution is sensitive to $a_\ast $ and $n$. Thus
for very small BH spin and value of $n$, the MC effects may be
ignored, and the solutions return to NDAF solutions. Considering
that the BH is spun-down in the MC process, while it is spun-up in
the accretion process. The two processes with opposite effects
result in a state with an equilibrium spin $a_\ast ^{eq} $.
Calculation shows that $a_\ast ^{eq} $ is greater than 0.85 for $n >
3$. This result implies that the MC effects are dominant in the
whole duration of GRB, for which a fast-spinning BH is the central
engine.

Secondly, we made many simplifications in the MCNDAF model, such as
we omit the photodisintegration term in the cooling rates, we assume
the disk is thin, and a very simple magnetic configuration, and so
on. Recently, Liu et al. (2007) took into account more realistic
microphysics. Chen {\&} Beloborodov (2006) worked out the NDAF
solution under full Kerr metric. It is necessary to combine these
effects with the MC process and work out a more detailed model in
the future.

Finally, our MCNDAF is steady. Recently, Janiuk et al. (2004, 2007) computed
the time evolution of NDAF that proceeds during the burst. It is very
interesting to investigate a time-dependent MCNDAF.

Recently, Zhang {\&} Dai (2007) proposed a hyperaccretion disk
around a neutron star. They found, compared with a BH disk, the
hyperaccretion disk around a neutron star can be cooled more
efficiently and produce a much higher neutrino luminosity. As
discussed by Kluzniak {\&} Rappaport (2007), the magnetic dipole of
the neutron star can also torque the disk. Therefore, it is also
attractive to consider the effects of the magnetic torque in the
context of hyperaccretion disk around a neutron star.

\acknowledgments
We thank T. Liu for helpful discussions, and also
thank the anonymous referee for his/her valuable comments and
constructive suggestions". This work is supported by National
Natural Science Foundation of China under Grants 10873005, 10847127
and 10703002, the Research Fund for the Doctoral Program of Higher
Education under grant 200804870050 and National Basic Research
Program of China under Grant No. 2009CB824800.

\clearpage
\begin{figure}[htbp]
\plotone{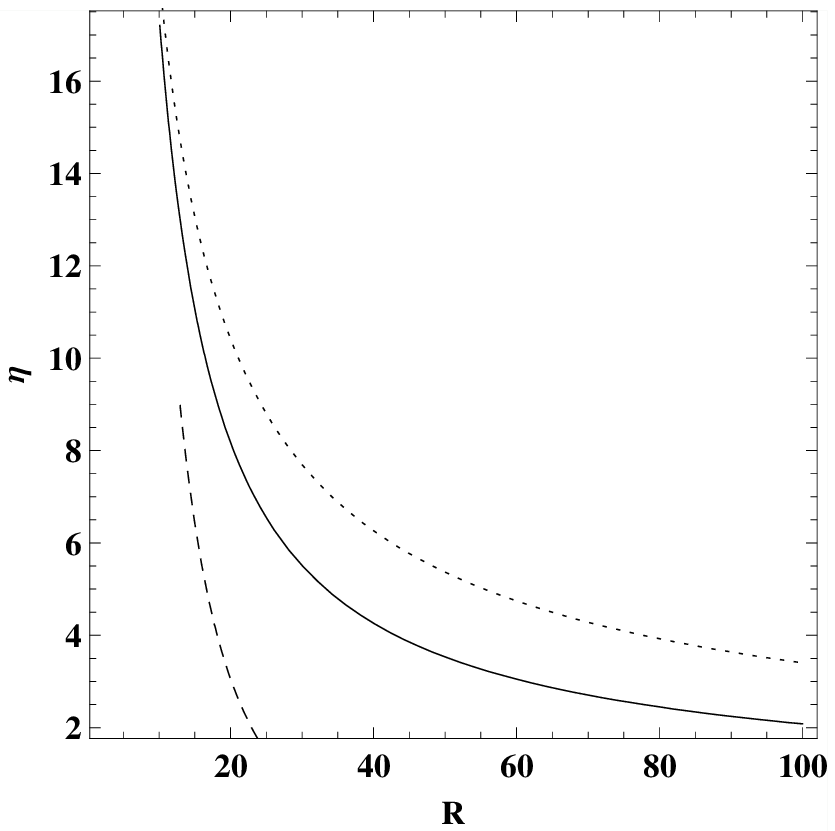} \label{fig1} \caption{The curves of the ratio of
$Q_{MC} $ to $Q_G $ versus the disk radius R for $a_\ast = 0.9$ with
$n=3$ (solid line), $a_\ast = 0.9$ with $n=4$ (dotted line) and
$a_\ast = 0.8$ with $n=3$ (dashed line).}
\end{figure}

\clearpage
\begin{figure}[htbp]
\plotone{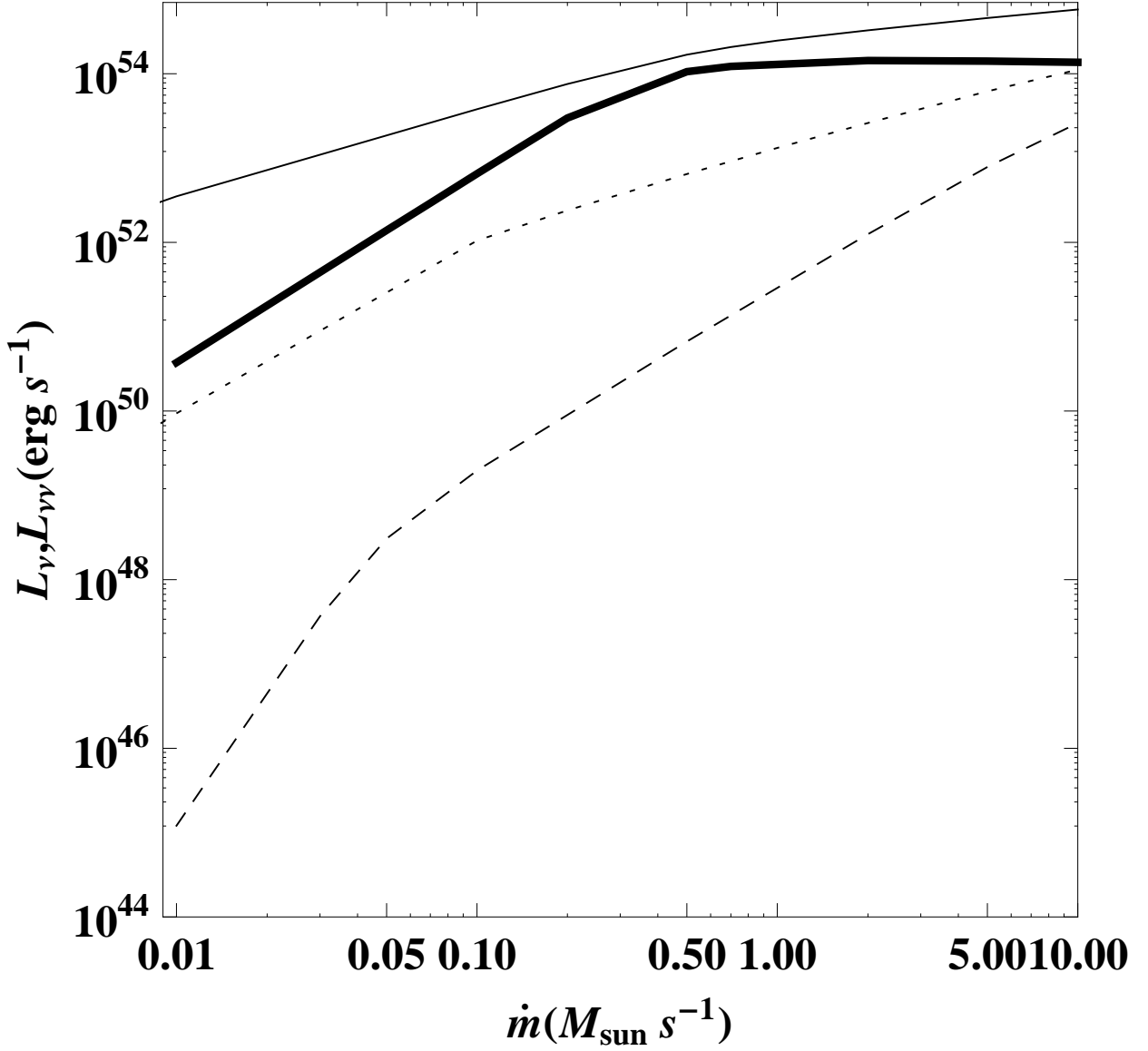} \label{fig2} \caption{The total neutrino
luminosity $L_\nu $ and $L_{\nu \bar {\nu }} $ for $a_\ast = 0.9$
and $n=3$. The thick and thin solid lines represent $L_{\nu \bar
{\nu }} $ and $L_\nu $ of MCNDAF, respectively. The dashed line and
dotted line represent $L_{\nu \bar {\nu }} $ and $L_\nu $ of NDAF
without MC, respectively.}
\end{figure}

\clearpage
\begin{figure}[htbp]
\plotone{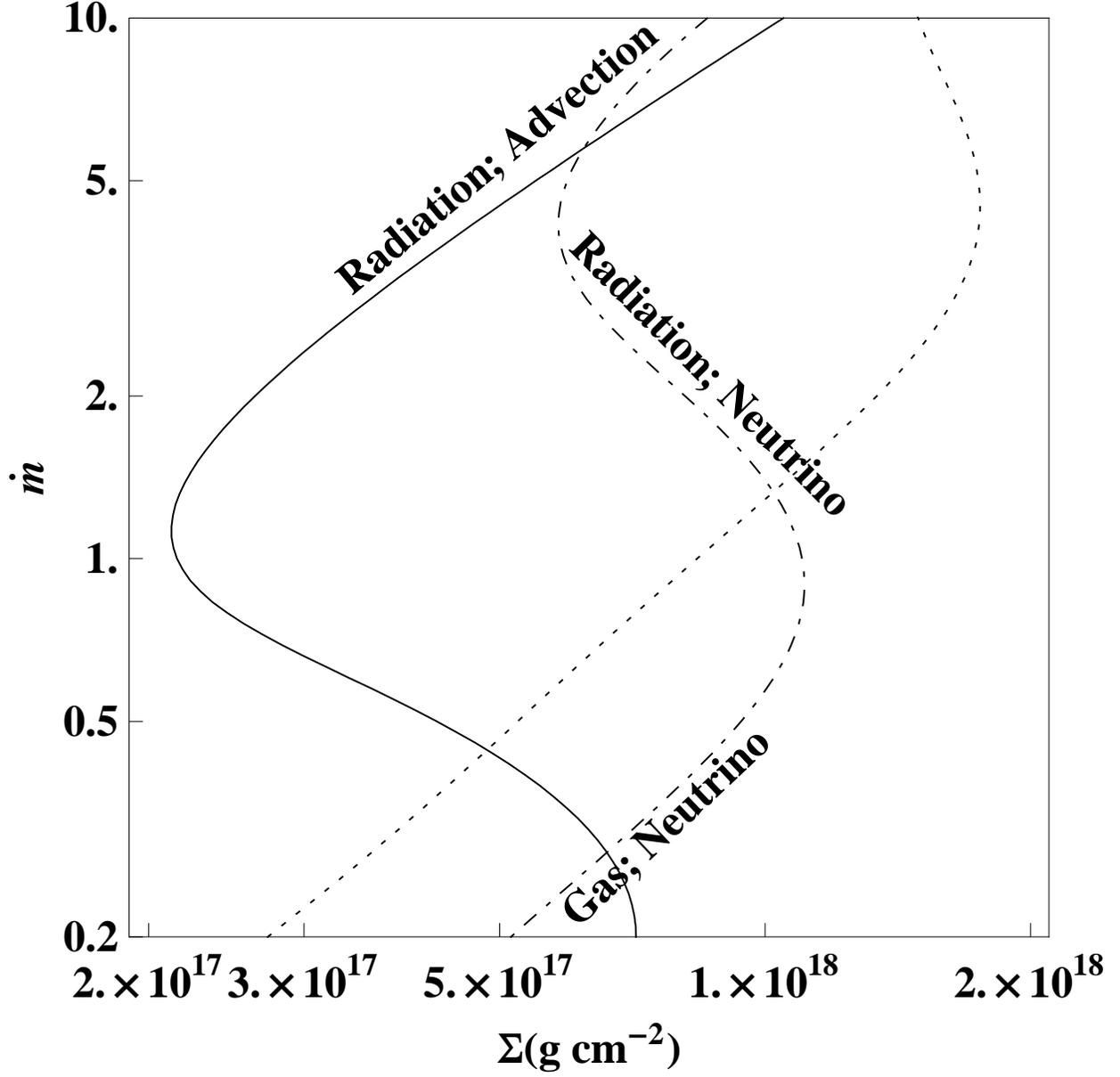} \label{fig3} \caption{The curves of $\dot {m}$
versus $\Sigma $ for several given disk radii with $R=5, 10$ and
20 in solid, dashed and dotted lines, respectively. Other parameters are $a_\ast = 0.9$
and $n=3$.}
\end{figure}

\clearpage
\begin{figure}[htbp]
\plottwo{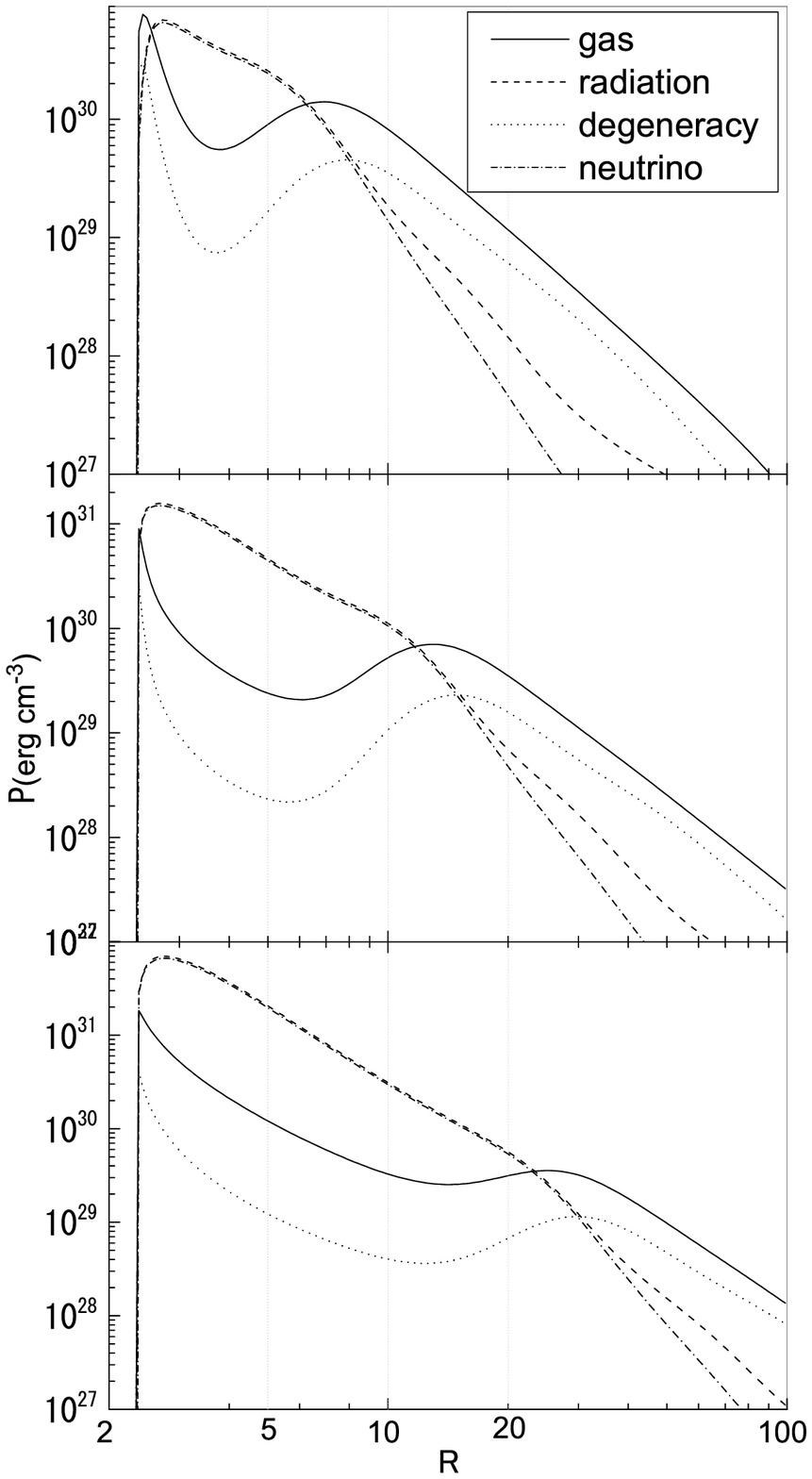}{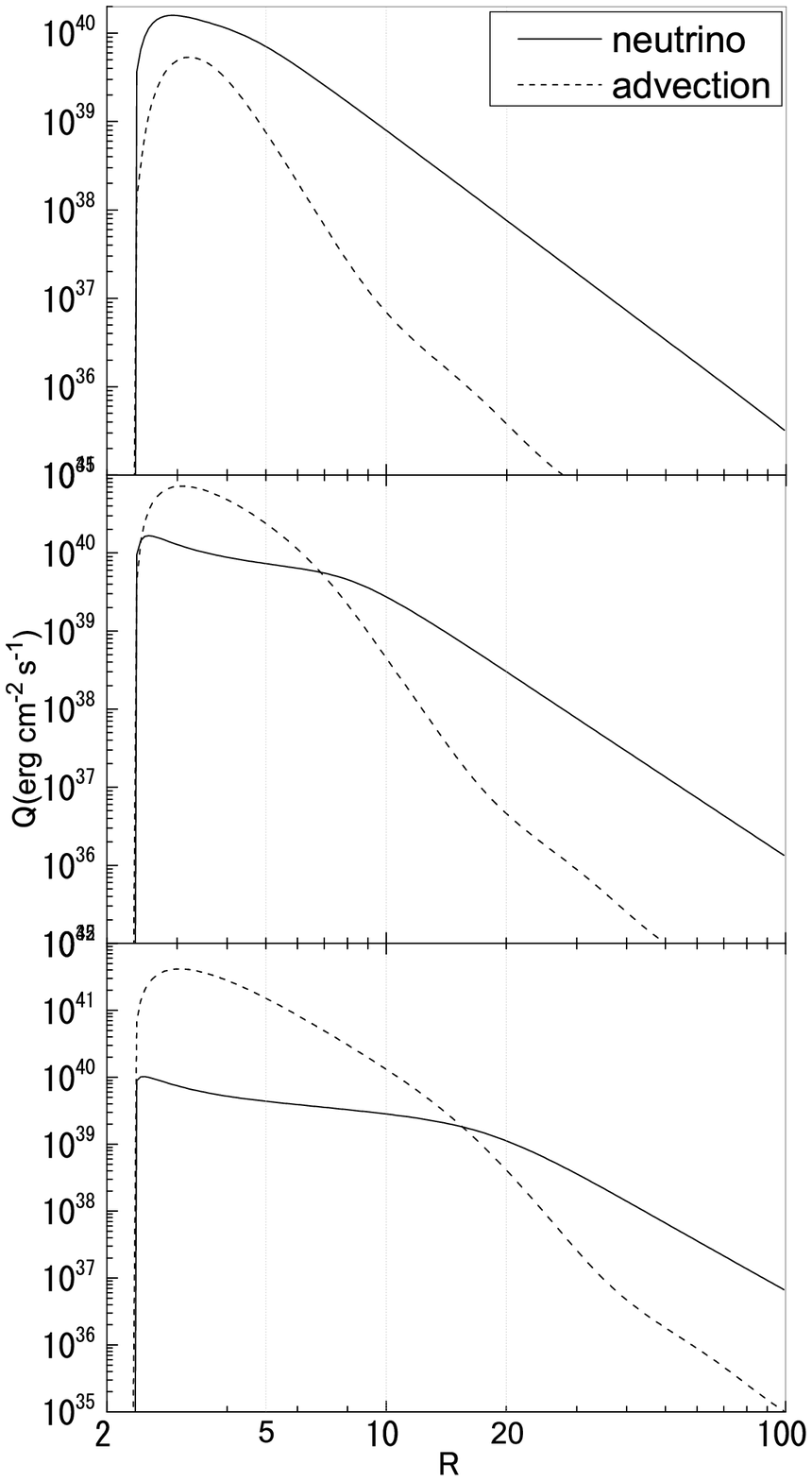} \label{fig4} \caption{Pressure (left)
and cooling rate (right) components as a function of the disk
radius, for three accretion rate values: $\dot {m} = 0.5$ (top),
$\dot {m} = 2$ (middle) and $\dot {m} = 10$ (bottom). The pressure
components are: gas pressure (solid line), radiation pressure
(dashed line), degeneracy pressure (dotted line), and neutrino
pressure (dot-dashed line). The cooling terms are: cooling rates due
to neutrino emission (solid line) and advection (dashed line). Other parameters are $a_\ast = 0.9$
and $n=3$.}
\end{figure}

\clearpage
\begin{figure}[htbp]
\plotone{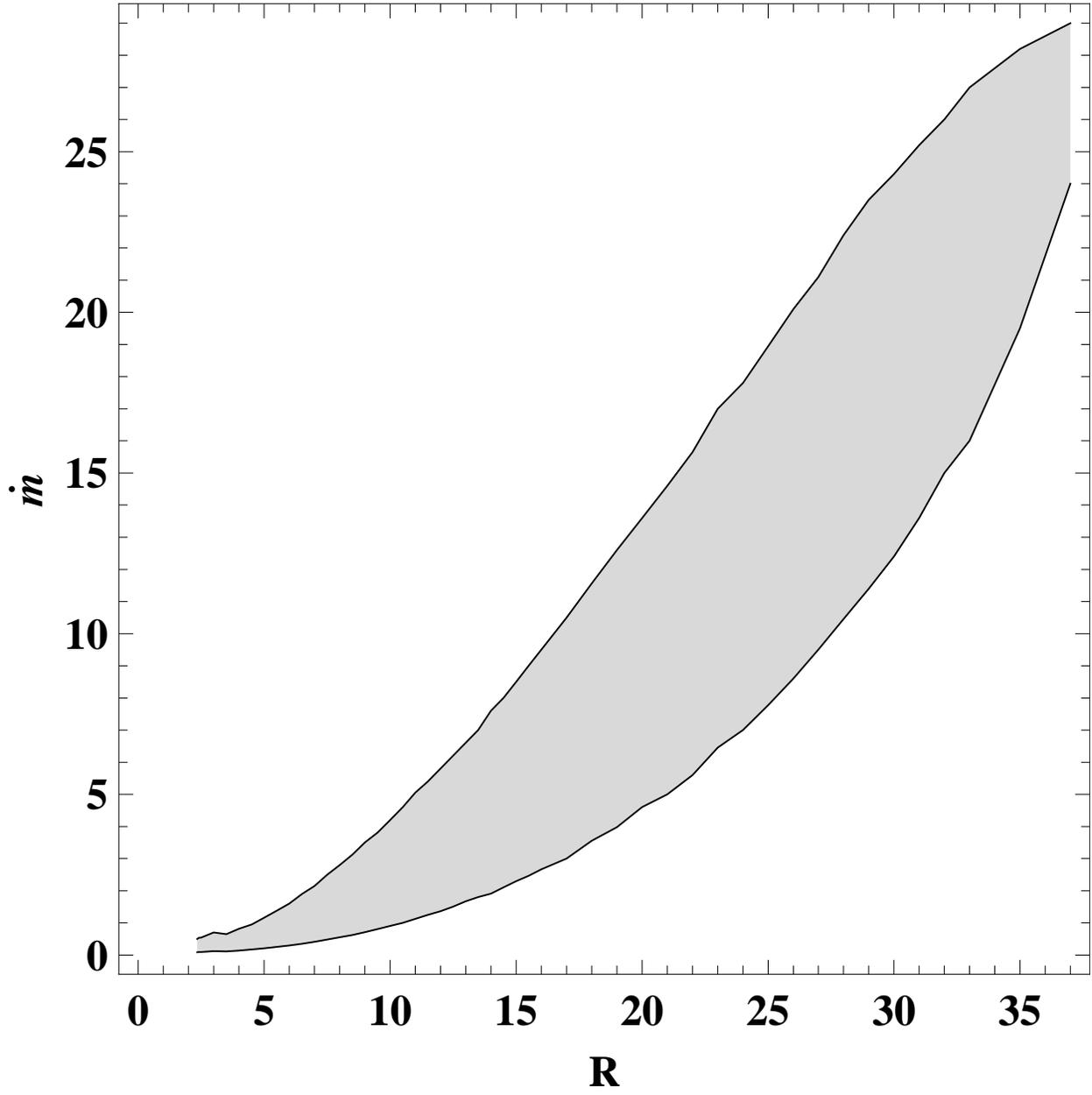} \label{fig5} \caption{The viscous instability is
indicated by the shaded region in the parameter space. The parameters are $a_\ast = 0.9$
and $n=3$.}
\end{figure}

\end{document}